\documentclass[12pt,a4paper]{article}

\usepackage{amsmath,amsfonts,amssymb,amsthm}
\usepackage{tikz}  
\usetikzlibrary{snakes}
\usepackage{cite}

\renewcommand{\a}{\alpha}           

\newcommand{\del}{\partial}         
\newcommand{\dl}{\delta}            
\newcommand{\g}{\mathfrak{g}}       
\newcommand{\ga}{\gamma}            
\newcommand{\gu}{\mathfrak{u}}      
\renewcommand{\L}{\mathcal{L}}      
\newcommand{\nn}{\nonumber}         
\newcommand{\ovl}{\overline}        
\newcommand{\owl}{\overline}        
\newcommand{\Sf}{\mathbb{S}}        
\newcommand{\sepword}[1]{\quad\hbox{#1}\quad} 
\DeclareMathOperator{\T}{T}         
\newcommand{\thalf}{\tfrac{1}{2}}   
\newcommand{\thW}{\theta_{\mathrm{W}}} 
\newcommand{\tquarter}{\tfrac{1}{4}} 
\newcommand{\ut}{{\tilde u}}        
\newcommand{\vf}{\varphi}           
\newcommand{\x}{\times}             
\renewcommand{\.}{\cdot}            
\def\wick:#1:{\mathopen:#1\mathclose:} 
\def\<#1,#2>{\langle#1,#2\rangle}   

\theoremstyle{plain}

\theoremstyle{definition}

\begin{document}

\title{On the causal gauge principle}

\author{
Jos\'e M. Gracia-Bond\'{\i}a\,\dag\,\ddag\
\\
\dag Departamento de F\'{\i}sica Te\'orica,
\\
Universidad de Zaragoza, Zaragoza 50009, Spain
\\
and
\\
\ddag Departamento de F\'{\i}sica,
\\
Universidad de Costa Rica, San Pedro 2060, Costa Rica
}

\date{\today}

\maketitle

\begin{abstract}
\smallskip Work by the Z\"urich school of causal (Epstein--Glaser)
renormalization has shown that renormalizability in the presence of
massless or massive gauge fields (as primary entities) explains gauge
invariance and, in some instances, the presence of a Higgs-like
particle, without need for a
Brout--Englert--Higgs--Guralnik--Hagen--Kibble (BEHGHK) mechanism. We
review that work, in a pedagogical vein, with a pointer to go~beyond.
\end{abstract}

\tableofcontents

\section{Introduction}

By now spontaneous symmetry breaking (SSB) of local symmetry is a
well-established paradigm of high-energy physics.  At the end of the
60s and beginning of the 70s, it allowed the incorporation of
(electro)weak interactions into the framework of renormalizable field
theory.  In connection with the contemporaneous rise of the Standard
Model (SM), it enjoys immense historical success.

However, allusion to unsatisfactory or mysterious aspects of the Higgs
sector of the SM does pop up in the literature ---see for
instance~\cite[Sect.~22.10]{AH04}.  The Higgs self-coupling terms are
completely ad-hoc, unrelated to other aspects of the theory, and do
not seem to constitute a gauge interaction.  Moreover they raise the
hierarchy problem~\cite[Ch.~11]{ModernosPragmaticos}.  The most
frequent interpretation of the BEHGHK mechanism clashes with
cosmology~\cite{SolDedo}.

Debate on the proper interpretation of the mechanism (whether the
symmetry is ``broken'' or just ``hidden'', whether the Higgs field
truly has a non-zero vacuum expectation value (VEV) or
not~\cite{LosChinos}, and so on) seems endless.  This breds some
skepticism, even among earlier and doughty practitioners.  At the end
of his Nobel lecture~\cite{TiniDixit}, Veltman chose to declare:
\textit{``While theoretically the use of spontaneous symmetry
breakdown leads to renormalizable Lagrangians, the question of whether
this is really what happens in Nature is entirely open''.}

Indeed, since the \textit{deus ex machina} fields involved in broken
or hidden symmetry are unobservable, the status question for the
BEHGHK contraption cannot be resolved by the likely sighting of the
Higgs particle in the~LHC.%
\footnote{This situation has recently called the attention of
knowledgeable philosophers of
science~\cite{HombreOreja,NoClothesKing}: in epistemological terms,
they argue that the mechanism had heuristic value in the context of
discovery; but much less so in the context of justification.}

The subject has also been obscured all along by theoretical prejudice.
In the SM the Higgs field carries the load of giving masses to
\textit{all} matter and force fields.  For instance, it is said that
mass terms for the vector bosons are incompatible with gauge
invariance.  It ain't so: such mass terms fit in gauge theory by use
of St\"uckelberg fields~\cite{Altabonazo,Felicitas}.

Skepticism would be idle, nevertheless, in the absence of alternative
theoretical frameworks.  Assuming an agnostic stance, we pose the
question: is it possible to formulate the main results of
flavourdynamics, and to frame suggestions of new physics, without
recourse to unobservable processes?  In tune with the phenomenological
SM Lagrangian~\cite{SlimKilian}, this amounts to regard massive vector
bosons (MVB) as fundamental entities.

So let us stop pretending we know the origins of mass.  Higgs-like
scalar fields will still come in handy for either renormalizability or
unitarity; however, their gauge variations need not be the
conventional ones.  Fermions can be assigned Dirac masses, and
couplings with the scalar field proportional to those; this
contradicts in no way the chiral nature of their interactions in
the~SM.

An approach with the mentioned traits is already found in the
literature in the work by Scharf, D\"utsch and others, under the label
of the ``quantum gauge invariance'' principle.  A few references to it
are~\cite{PGI-EW-I,PGI-EW-II, CabezondelaSal, PepinsFriend} and mainly
the book~\cite{Zurichneverdies}.  The ``quantum Noether principle''
of~\cite{HurthS1,HurthS2} coincides essentially with it.  Both are
based on the rigorous causal scheme for
renormalization~\cite{PastMasters} by Epstein and Glaser (EG).

\smallskip

Henceforth we refer to the approach as causal gauge invariance (CGI).
The usual plan of the article is found at the end of the next section,
when the stakes hopefully have been made clearer.

\section{Overview of the CGI method}

The spirit of CGI is very much that of the~\cite{LuisDixit}.  Let $s$
denote the nilpotent BRS operation.  To realize gauge symmetry, one
should incorporate BRS symmetry ab initio in a ``quantum'' Lagrangian
$\L$, such that (very roughly speaking) $s\L\sim0$, and proceed to
build from there.  We do this for MVBs.

The starting point for the analysis is the Bogoliubov--Epstein--Glaser
functional scattering matrix on Fock space, in the form of a power
series:
\begin{equation}
\Sf(g) = 1 + \sum_{n=1}^\infty\frac{i^n}{n!}\,\int dx_1\ldots dx_n\,
T_n(x_1, \ldots, x_n)\,g(x_1) \cdots g(x_n).
\label{eq:dance-with-her}
\end{equation}
The coupling constants of the model are replaced by test functions
---we wrote just one of them for simplicity.  The theory is then
constructed basically by using causality and Poincar\'e invariance to
recursively determine the form of the time-ordered products~$T_n$ from
the $T_m$ with $m<n$; in this sense the procedure is inverse to the
``cutting rules''.  Only those fields should appear in~$T_n$ that
already are present in~$T_1$.  The procedure yields a finite
perturbation theory without regularization; ultraviolet divergences
are avoided by proper definition of the $n$-point functions as
distributions.

(Ultimately one would be interested in the adiabatic limit
$g(x)\uparrow g$.  This is delicate, however, due to infrared
problems.  We look at the theory before that limit is taken.)

With the proviso that two forms of $T_n$ are equivalent if they differ
by $s$-cobound\-aries, CGI is formulated by the fact that $sT_n$ must
be a divergence.  Roughly speaking, we must have
\begin{align}
sT_n(x_1,\ldots,x_n) &= i\sum_{l=1}^{l=n}
\T\bigl[T_1(x_1),\ldots,\del_l\.Q(x_l),\ldots,T_1(x_n)\bigr],
\nn \\
&=: i\sum_{l=1}^{l=n}\del_l\.Q_n(x_1,\ldots,x_n).
\label{eq:gold-mine}
\end{align}
for vectors $Q_n$, called $Q$-vertices, with $\del_l$ denoting the
partial divergences with respect to the~$x_l$ coordinates and $\T$ a
time-ordering operator.  In this way renormalization and gauge
invariance are linked in the EG scheme.  (We said ``roughly''
because~\eqref{eq:gold-mine} suggests that $\T$ and spacetime
derivatives commute, which is not generally the case for on-shell
fields.)

Note that $T_1$ only contains the first-order part of the Lagrangian.
Nevertheless, already the first order condition
\begin{equation*}
sT_1 = i\del\.Q_1 
\end{equation*}
constrains significantly the form of the Lagrangian.  Later on, we
show leisurely how the CGI method works for tree graphs belonging to
$T_2$.  This is almost all what is required for the purposes of this
paper: for ordinary gauge theories, the treatment of~$T_3$ is pretty
simple, and higher orders not needed at all.

Keep in mind that one works here with \textit{free} fields.
Interacting fields can be arrived at in the Epstein--Glaser procedure,
somewhat a posteriori, using their definition by Bogoliubov as
logarithmic functional derivatives of~$\Sf(g)$ with respect to
appropriate sources.  Their gauge variations resemble more those of
standard treatments; but we do not use them.  Thus $s$ ``sees'' only
the (massive or massless) gauge fields, and the attending (anti-)ghost
and St\"uckelberg fields.  This is why everything flows from the
quantum gauge structure of the boson sector.  Coupling to fermions,
which ought not be organized in multiplets a priori, comes almost like
an afterthought.

As it turns out, the procedure is quite restrictive, and in particular
only a few models for MVB theory pass muster.  These exhibit very
definite mass and interaction patterns, in particular quartic
self-interaction for the scalar particles.

We next compile the results, according to~\cite{Zurichneverdies}.
Consider a model with $t$ intermediate vector bosons $A_a$ in all, of
which any may be in principle massive or massless. Let us say there
are $r$ massive ones with masses $m_a, 1\le a\le r$ and $s$ massless
ones, and $t =r+s$. We assume there is \textit{one} (at most) physical
scalar particle~$H$ of mass $m_H$: \textit{entia non sunt
multiplicanda praeter necessitatem}. The BRS extension of the Wigner
representation theory for MVBs requires St\"uckelberg
fields~$B_a$~\cite{CabezondelaSal}, beyond the fermionic ghosts
$u_a,\ut_a$; in case $A_a$ is massless, we of course let $B_a$ drop
out. Adopting the Feynman gauge, the gauge variations are as follows:
\begin{align}
sA_a^\mu(x)   &= i\del^\mu u_a(x);
\nonumber \\
sB_a(x)       &= im_au_a(x);
\nonumber \\
su_a(x)       &= 0;
\nonumber \\
s\ut_a(x)     &= -i\big(\del\.A_a(x) + m_aB_a(x)\big).
\nonumber \\
sH(x)         &= 0.
\label{eq:begging-for-trouble}
\end{align}
This operator is nilpotent on-shell.

The total bosonic interaction Lagrangian, in a notation close to that
of~\cite{Zurichneverdies}, is of the form
\begin{equation}
\L_{\mathrm{int}} = gT_1 + \frac{g^2\,T_2}2,
\label{eq:romeros-somos}
\end{equation}
where $g$ is an overall dimensionless coupling constant;
$$
T_1 = f_{abc}\big(T^1_{1abc} + T^2_{1abc} + T^3_{1abc} +
T^4_{1abc}\big) + C\big(T^5_1 + T^6_1 + T^7_1 + T^8_1 + T^9_1\big)
$$
includes the cubic couplings, and
$$
T_2 = T^1_2 + T^2_2 + T^3_2 + T^4_2 + T^5_2 + T^6_2 + T^7_2
$$
includes the quartic ones.  The list of cubic couplings not
involving~$H$ is given by:
\begin{align}
T^1_{1abc} &= \bigl[A_a\.(A_b\.\del)A_c - u_b(A_a\.\del\ut_c)\bigr];
\nn \\
T^2_{1abc} &= \frac{m_b^2 + m_c^2 - m_a^2}{4m_bm_c}
\bigl[B_b(A_a\.\del B_c) - B_c(A_a\.\del B_b)\bigr];
\nn \\
T^3_{1abc} &= \frac{m_b^2 - m_a^2}{2m_c}(A_a\.A_b)B_c; 
\nn \\
T^4_{1abc} &= \frac{m_a^2 + m_c^2 - m_b^2}{2m_c}\ut_au_bB_c;
\label{eq:abyssus-abyssum-invocat}
\end{align}
The list of cubic couplings of the Higgs-like particle is:
\begin{align}
T^5_1 &= m_a[B_a(A_a\.\del H) - H(A_a\.\del B_a)];
\nonumber \\
T^6_1 &= m_a^2(A_a\.A_a) H; 
\nonumber \\
T^7_1 &= -m_a^2\ut_au_a H;
\nonumber \\
T^8_1 &= -\thalf m_H^2 B_a^2 H;
\nonumber \\
T^9_1 &= -\thalf m_H^2 H^3.
\label{eq:dente-superbo}
\end{align}
Remarks: in~\eqref{eq:abyssus-abyssum-invocat}
and~\eqref{eq:dente-superbo} we sum over repeated indices; the
$f_{abc}$ are completely skewsymmetric in their three indices, and
fulfil the Jacobi identity; $T^1_{1.}$ yields the cubic part in the
classical Yang--Mills Lagrangian; $C$ is a constant independent
of~$a$.  The dimension of the Lagrangian must be $M^4$ in natural
units, and the boson field dimension in our formulation is~1 for
\textit{both} spins: the dimension of $C$ is $M^{-1}$.  Note the
diagonality of the couplings of the Higgs-like particle.  Crossed
terms like $(A_a\.A_b)H$ for $a\ne b$, and others like
$B_aB_bB_c,B_aH^2\ldots$, that could be envisaged, are held to vanish
by CGI.

The list of quartic couplings:
\begin{align}
T^1_2 &= -\thalf f_{abc}f_{ade}(A_b\.A_d)(A_c\.A_e);
\nn \\
T^2_2 &= \bigg[\frac{(m_d^2 + m_e^2 - m_a^2)(m_c^2 + m_e^2 - m_b^2)}
{8m_dm_cm_e^2}f_{ade}f_{bce} + c \leftrightarrow d
\nn \\
&+ \thalf C^2m_am_b\dl_{ad}\dl_{bc} + c \leftrightarrow d \bigg]
\x (A_a\.A_b)B_cB_d;
\nn \\
T^3_2 &= -\tquarter C^2 m_H^2 B^2_aB^2_b \sepword{irrespective of 
$a,b\le r$;}
\nn \\
T^4_2 &= Cf_{abc}\frac{m_b^2 - m_a^2}{m_c}(A_a\.A_b)B_c H;
\nn \\
T^5_2 &= C^2m_a^2(A_a\.A_a)H^2;
\nn \\
T^6_2 &= -\thalf C^2 m_H^2 B^2_a H^2 \sepword{irrespective of $a\le
r$;}
\nn \\
T^7_2 &= -\tquarter C^2 m_H^2 H^4.
\label{eq:hoc-erat-in-votis}
\end{align}

Every coefficient of the interaction Lagrangian is in principle
determined in terms of the $f_{abc}$ and the pattern of masses. We are
not through, because CGI implies \textit{constraints}, in general
non-linear and extremely restrictive, on \textit{allowed patterns} of
masses for the gauge fields. But we may anticipate a few more
comments. The first term $T^1_2$ in~\eqref{eq:hoc-erat-in-votis} just
yields the quartic part in the classical Yang--Mills Lagrangian, as
expected. In case all the $A_a$ are massless, there is no need to add
physical or unphysical scalar fields for renormalizability, and only
$T^1_1$ and $T^1_2$ survive in the theory; they of course coincide
respectively with the first and second order part of the usual
Yang--Mills Lagrangian. In particular, CGI gives rise to gluodynamics.
(It must be said, though, that the physical equivalence of couplings
differing in a divergence is less compelling in this case, since there
is no asymptotic limit for the Bogoliubov--Epstein--Glaser
$\Sf(g)$-matrix; CGI offers no tools to deal with this infrared
problem.) Remarkably, with independence of the masses, CGI
unambiguously leads to generalized Yang--Mills theories on reductive
Lie algebras; apparently this was realized first by
Stora~\cite{ESItalk}.

\smallskip

The plan of the rest of the article is as follows.  Notice that the
case $r = 1,s = 0$ leads to an abelian model in which all the terms
with the Higgs-like field~$H$ survive.  We use this example in
Section~3 to illustrate in some detail ---missing
in~\cite{Zurichneverdies}--- how the second-order condition determines
the couplings.  Section~4 deals with \textit{three} gauge fields
---there are no models with two gauge fields to speak of, since
$\gu(1)\oplus\gu(1)$ is the only two-dimensional reductive Lie
algebra.  For that we need to invoke the mentioned mass relations
(reference~\cite{Zurichneverdies} unfortunately contains misprints in
this respect).  Section~5 elaborates on the reconstruction of the SM
in CGI, looking at the fermion sector as well.  The paper ends with a
discussion.

\section{The abelian model}
 
Consider a theory with a neutral gauge field~$A$ of mass~$m$ and a
\textit{physical} neutral scalar field~$H$ of mass~$m_H$, and basic
coupling~$AAH$.

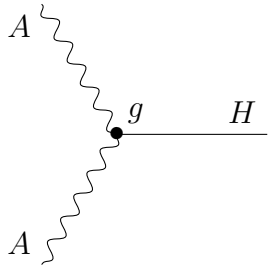
\begin{figure}[htb]
\centering
\begin{tikzpicture}
\draw (0,0) node {$\bullet$} node[above right] {$g$}
   -- (2,0) node [above left] {$H$};
\draw[snake=snake] (0,0) -- (120:2) node[below left] {$A$};
\draw[snake=snake] (0,0) -- (240:2) node[above left] {$A$};
\end{tikzpicture}
\end{figure}

Since massive quantum electrodynamics is known to be renormalizable
without an extra scalar field, this is perhaps not very interesting;
but our aim here is merely showing the workings of the causal gauge
principle.

\subsection{The first-order analysis}

For $T_1$, take the most general Ansatz containing cubic terms in the
fields and leading to a renormalizable theory.  With the benefit of
hindsight, we write down on the first line the terms destined to
survive:
\begin{align}
T_1/m &= (A\.  A)H + b\ut uH + c\big(H(A\.\del B) - B(A\.\del H)\big) +
dB^2H + eH^3
\nn \\
&+ a(A\.  A)B + b_2\ut uB + b_3u(A\.\del\ut) + d_1B^3 + d_3BH^2.
\label{eq:few-are-the-chosen}
\end{align}
The factor $m$ is natural according to our previous discussion on
dimensions.  The symmetric combination $HA\.\del B+BA\.\del H$ has
been excluded for the following reason:
$$
A\.(B\del H + H\del B) = \del.(BHA) - (\del\.A)BH,
$$
and in view of~\eqref{eq:begging-for-trouble}, the $(\del\. A)BH$
term is $s$-exact apart from terms of already present
in~\eqref{eq:few-are-the-chosen}.  Concretely,
$$
s(\ut BH) = -(\del\. A)BH - mB^2H - m\ut uH.
$$

We calculate next $sT_1/m$ in~\eqref{eq:few-are-the-chosen} and obtain 
for the first group of terms:
\begin{align}
&2\del\.(uHA) - 2u(\del\.  A)H - 2uA\.\del H - bu(\del\.A)H
\nn \\
&- bmu BH + c\del\. u(H\del B - B\del H)
\nn \\
& + cm[HA\.(\del u) - u A\.\del H] + 2dmuBH.
\label{eq:troppo-lavoro}
\end{align}
We have used $s(uC)=-usC$ for any~$C$.  In detail:
$$
-is(A\.AH) = (\del u\.A)H = 2\big[\del\.(uHA) - u(\del\.A)H - u
A\.\del\. H].
$$
Next
$$
-is(\ut uH) = -u(\del\.A)H - muBH.
$$
Next
$$
-is(A\.(H\del B - B\del H)) = \del u\.(H\del B - B\del H)
+ m[HA\.(\del u) - uA\.\del H].
$$
Finally $-is(B^2H)=2muBH$.

Similarly, for the second group of terms we obtain:
\begin{align*}
&2a\del\.(uBA) - 2au(\del\.A)B
- 2auA\.\del B + amuA\.A
\nn \\
& - b_2u(\del.A)B - b_2mu B^2 + b_3\big(\del u\.u\,
\del\ut + uA\.\del(\del\.A + mB)\big)
\nn \\
&+ 3d_1muB^2 + d_3muH^2.
\end{align*}
All terms of that group are excluded because their contributions
to~$sT_1$ are not pure divergences.  For instance, the first one
corresponds to the term in $uA\.A$, that can be canceled only by
setting $a=0$.

On the other hand, the second term in the second line
in~\eqref{eq:troppo-lavoro} can be recast as
$$
\del\.\bigl(u(H\del B - B\del H)\bigr) + (m^2 - m_H^2)uBH.
$$
For the following terms we have
$$
A\.(\del u)H - uA\.\del H = \del\.(uHA)
- u(\del\.A)H - 2u A\.\del H.
$$
In all, 
\begin{align*}
&-isT_1/m = \del\.(C + D) - (2 + cm + b)u(\del\. A)H
\\
&- (2 + 2cm)u(A\.\del H) + \big(2dm - bm + c(m^2 - m_H^2)\big)uBH;
\end{align*}
with the vectors $C,D$ given by $C:=(2+cm)uHA;\,D:=cu(H\del B-B\del
H)$.  The terms that are not a divergence must cancel.  This at once
leads to:
$$
c = -\frac{1}{m}; \; b = -1; \; d = -\frac{m_H^2}{2m^2}; \sepword{thus}
C = uHA; \quad D=\frac{-u}{m}(H\del B - B\del H).
$$

In summary, we have obtained the cubic couplings in the Lagrangian:

\begin{figure}[htb]
\begin{tikzpicture}
\centering
\draw (0,0) node {$\bullet$} node[above right] {$gm$}
   -- (2,0) node [above left] {$H$};
\draw[snake=snake] (0,0) -- (120:2) node[below left] {$A$};
\draw[snake=snake] (0,0) -- (240:2) node[above left] {$A$};
\end{tikzpicture}
\hspace{2pc}
\begin{tikzpicture}
\draw (0,0) node {$\bullet$} node[above right] {$-gm$} -- (2,0)
  node [above left] {$H$};
\draw[dashed] (0,0) -- (120:2) node[below left] {$u$};
\draw[dashed] (0,0) -- (240:2) node[above left] {$\tilde u$};
\end{tikzpicture}
\hspace{2pc}
\begin{tikzpicture}
\draw (0,0) node {$\bullet$} node[above right] {$\frac{-gm_H^2}{2m}$}
   -- (2,0) node [above left] {$H$};
\draw[dashed] (0,0) -- (120:2) node[below left] {$B$};
\draw[dashed] (0,0) -- (240:2) node[above left] {$B$};
\end{tikzpicture}
\end{figure}

\smallskip

\begin{figure}[htb]
\centering
\begin{tikzpicture}
\draw (0,0) node {$\bullet$} node[above right] {$g$}
   -- (2,0) node [above left] {$\del H$};
\draw[snake=snake] (0,0) -- (120:2) node[below left] {$A$};
\draw[dashed] (0,0) -- (240:2) node[above left] {$B$};
\end{tikzpicture}
\hspace{4pc}
\begin{tikzpicture}
\draw[dashed] (0,0) node {$\bullet$} node[above right] {$-g$}
   -- (2,0) node [above left] {$\del B$};
\draw[snake=snake] (0,0) -- (120:2) node[below left] {$A$};
\draw (0,0) -- (240:2) node[above left] {$H$};
\end{tikzpicture}
\caption{cubic vertices}
\end{figure}
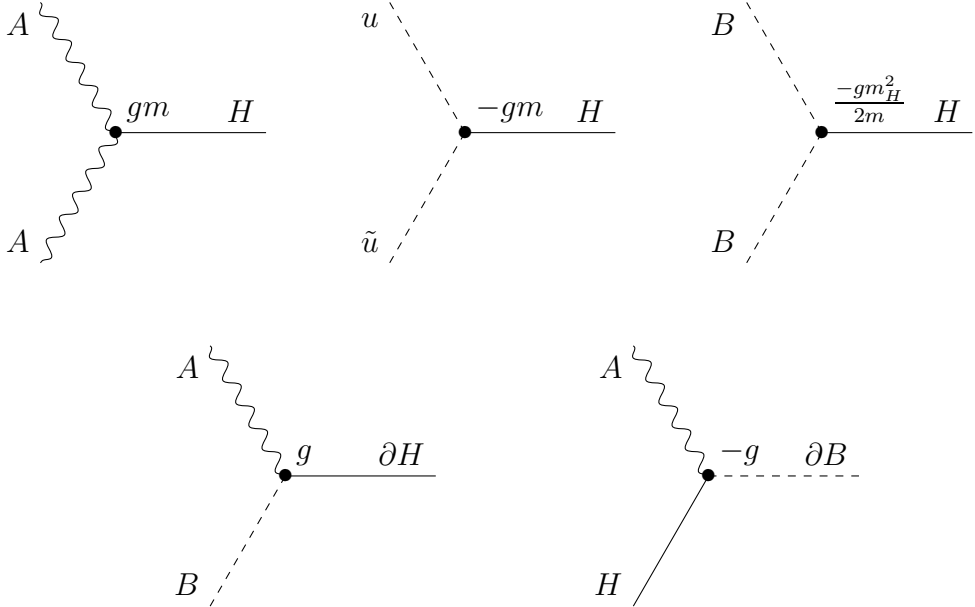

To this we should add the $H^3$ coupling, whose coefficient is still
indeterminate:

\newpage
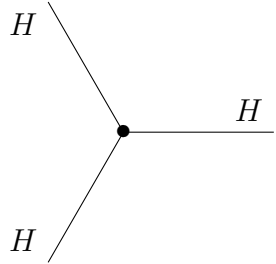
\begin{figure}[htb]
\centering
\begin{tikzpicture}
\draw (0,0) node {$\bullet$} node[above right] {}
   -- (2,0) node [above left] {$H$};
\draw (0,0) -- (120:2) node[below left] {$H$};
\draw (0,0) -- (240:2) node[above left] {$H$};
\end{tikzpicture}
\end{figure}
Moreover:
\begin{equation}
sT_1 = i\del\.Q_1 \sepword{with} Q_1 = muHA - u(H\del B - B\del H).
\label{eq:la-madre-del-cordero}
\end{equation}

\subsection{The second-order analysis}
 
The next step is less trivial. Equation~\eqref{eq:gold-mine} 
certainly makes sense outside the diagonals, for then the $\T$ 
product is calculated like an ordinary product. But the extension to 
the diagonals, which is simply $x_1=x_2$ for $n=2$, can produce local 
correction terms.  At this order, the advanced and retarded products 
are given by:
\begin{align}
A_2(x_1,x_2) &= T_2(x_1,x_2) - T_1(x_1)T_1(x_2);
\nonumber \\
R_2(x_1,x_2) &= T_2(x_1,x_2) - T_1(x_2)T_1(x_1);
\label{eq:sleeping-dog}
\end{align}
Here $T_2(x_1,x_2)$ is still unknown, but it is clear that $A_2$ will
have support on the past light cone of~$x_2$, and $R_2$ on its future
light cone; hence the nomenclature.  Consider then $D_2(x,y):=
\big(R_2-A_2\big)(x,y)=[T_1(x),T_1(y)]$, whose support is within the light
cone (we say $D_2$ is causal).  We have thus
\begin{align}
sD_2(x,y) &= [sT_1(x), T_1(y)] + [T_1(x), sT_1(y)]
\nonumber \\
&= i\del_x[Q_1(x), T_1(y)] + i\del_y[T_1(x), Q_1(y)];
\label{eq:salta-la-liebre}
\end{align}
so that $D_2$ moreover \textit{is} gauge-invariant.  The crucial step
in EG renormalization is the \textit{splitting} of $D_2$ into the
retarded part~$R_2$ and the advanced part~$A_2$; once this is done,
$T_2$ is found at once from~\eqref{eq:sleeping-dog}.  The issue is how to
preserve gauge invariance in this distribution splitting.  For this,
we split $D_2$ and the commutators ---without the derivatives--- in
the previous equation; then gauge invariance:
$$
sR_2(x,y) = i\del_x R_{2/1}(x,y) + i\del_y R_{2/2}(x,y)
$$
can only be (and is) violated for $x=y$, that is, by local terms
in~$\dl(x-y)$.  However, if in turn local renormalization terms
$N_2,N_{2/1},N_{2/2}$ can be found in such a way that
$$
s(R_2(x,y) + N_2(x,y)) = i\del_x(R_{2/1} + N_{2/1}) + i\del_y(R_{2/2} +
N_{2/2}),
$$
with an obvious notation, then CGI to second order holds. 

To the purpose we consider only tree diagrams.  In view
of~\eqref{eq:salta-la-liebre}, we systematically proceed to study the
divergences coming from cross-terms
between~\eqref{eq:la-madre-del-cordero} and
\begin{equation}
T_1 = m\bigl[(A\.A)H + u\ut H - \frac{1}{m}A\.\bigl(H\del B - B\del
H\bigr) - \frac{m_H^2}{2m^2}\,B^2H + eH^3\bigr].
\label{eq:donya-toda}
\end{equation}
Factors containing derivatives give rise to normalization
contributions after distribution splitting.

\smallskip

The most difficult part of the coming calculation asks for divergences
of terms with commutators $[\del^\mu B(x), \del^\nu B(y)]$ and
$[\del^\mu H(x), \del^\nu H(y)]$.  Following~\cite{Michael}, we look
at Section~4 in~\cite{PGI-EW-I} in order to prepare the computation.
There, for general functions $F,E$ we find the formulas:
\begin{align}
&\del^x_\mu[F(x)E(y)\dl(x - y)] + \del^y_\mu[F(y)E(x)\dl(x - y)]
\nonumber \\
&= \del_\mu F(x)\,E(x)\dl(x - y) + F(x)\,\del_\mu E(x)\dl(x - y)
\label{eq:villanous} \\
\sepword{and} & F(x)E(y)\del^x_\mu\dl(x - y) + F(y)E(x)\del^y_\mu\dl(x - y) 
\nonumber \\
&= F(x)\,\del_\mu E(x)\dl(x - y) - \del_\mu F(x)\,E(x)\dl(x - y).
\label{eq:amorphous}
\end{align}
We may prove both from the following observation: since
$$
F(x)E(y)\dl(x - y) = F(x)E(x)\dl(x - y),
$$
it must be that
$$
\del^x_\mu\big(F(x)E(y)\dl(x - y)\big) = \del^x_\mu\big(F(x)E(x)\dl(x - 
y)\big);
$$
which forces
\begin{equation}
E(y)\del^x_\mu\dl(x - y) = E(x)\del^x_\mu\dl(x - y) + \del_\mu
E(x)\dl(x - y).
\label{eq:penguins}
\end{equation}
Now,
\begin{align*}
&\del^x_\mu[F(x)E(y)\dl(x - y)]  + \del^y_\mu[F(y)E(x)\dl(x - y)]
\\
&= \del_\mu F(x)\,E(x)\dl(x - y) + F(x)E(y)\del^x_\mu\dl(x - y)
\\
&+ \del_\mu F(x)\,E(x)\dl(x - y)  - F(y)E(x)\del^x_\mu\dl(x - y) 
\\
&= \del_\mu F(x)\,E(x)\dl(x - y) + F(x)E(y)\del^x_\mu\dl(x - y)
\\
&- F(x)E(x)\del^x_\mu\dl(x - y) = \del_\mu F(x)\,E(x)\dl(x - y) +
F(x)\,\del_\mu E(x)\dl(x - y);
\end{align*}
where we have used~\eqref{eq:penguins} twice.  Analogously,
\begin{align*}
&F(x)E(y)\del^x_\mu\dl(x - y) + F(y)E(x)\del^y_\mu\dl(x - y)
= F(x)E(x)\del^x_\mu\dl(x - y)
\\
&+ F(x)\del_\mu E(x)\dl(x - y) - F(y)E(x)\del^x_\mu\dl(x - y)
\\
&= F(x)\,\del_\mu E(x)\dl(x - y) - \del_\mu F(x)\,E(x)\dl(x - y),
\end{align*}
using~\eqref{eq:penguins} twice again.

\smallskip

We finally start the advertised computation.  Coming from respectively
the second term of $Q_1(x)$ in~\eqref{eq:la-madre-del-cordero} and
third of~$T_1(y)$ in~\eqref{eq:donya-toda}, now we find for $i[Q_1(x),
T_1(y)]$:
\begin{align*}
&iu(x)H(x)[\del^\mu B(x), \del^\nu B(y)]A_\nu(y)H(y)
\\
&= u(x)H(x)A_\nu(y)H(y)\del^\mu_x\del^\nu_y D(x - y).
\end{align*}
The identity $[B(x), B(y)]=-iD(x-y)$ for scalar fields has been
employed.  Next we need to tackle the divergence of the splitting
of~$\del^\mu_x\del^\a_yD$.  Splitting of the Jordan--Pauli
propagator~$D$ gives rise to the retarded propagator~$D^{\rm ret}$.
Now, each derivation increases by one the singular order of a
distribution.  Thus, although $\del_x^\mu\del_y^\nu D^{\rm ret}$ is a
well-defined distribution, its singular order is~$-2+2=0$, therefore
allowing a normalization term in the split~distribution:
$$
\del_x^\mu\del_y^\nu D^{\rm ret}(x-y) \to \del_x^\mu\del_y^\nu D^{\rm
ret}(x - y) + C_Bg^{\mu\nu}\dl(x - y).
$$
After applying~$\del_\mu$, simply from
$$
\del^x_\mu\del_x^\mu D^{\rm ret}(x-y) = - m^2D^{\rm ret}(x - y) +
\dl(x - y),
$$
the total singular part is of the form
$$
C_B\del^\nu_x[F(x)E(y)\,\dl(x - y)] + F(x)E(y)\del^\nu_y\dl(x - y),
\sepword{with} F = uH; \; E = HA_\nu.
$$
Adding the term with $x$ and~$y$ interchanged, and using the
identities~\eqref{eq:villanous} and~\eqref{eq:amorphous}, it comes
finally the short rule for this kind of singular term:
$$
F(x)E(y)\del^\mu_x\del^\nu_yD(x - y) \to [(C_B + 1)(\del^\nu F)E +
(C_B - 1)F\,\del^\nu E]\dl(x - y).
$$
Therefore we obtain in the end
\begin{align}
&\quad (C_B + 1)\big[H^2(A\.\del u) + uH(A\.\del H)\big]\dl(x - y)
\label{eq:bitter-end}
\\
&+ (C_B - 1)\big[uH^2(\del\.A) + uH(A\.\del H)\big]\dl(x - y).
\nonumber
\end{align}

By the same token, coming now from respectively the third and fourth
terms in $Q_1(x)$ and~$T_1(y)$, and performing entirely similar
operations, we obtain
\begin{align}
&\quad (C_H + 1)\big[B^2(A\.\del u) + uB(A\.\del B)\big]\dl(x - y)
\label{eq:RIP}
\\
&+ (C_H - 1)\big[uB^2(\del\.A) + uB(A\.\del B)\big]\dl(x - y).
\nonumber
\end{align}
There is no good reason for $C_H\ne C_B$; see further on.

\smallskip

There are no singular contributions from the first term in~$Q_1(x)$.
The second term there will contribute for the commutators with the
fourth and fifth terms in~$T_1(y)$.  Concretely, there is the term
\begin{align*}
&-iu(x)H(x)[\del^\mu B(x), B(y)]A_\nu(y)\del^\nu H(y)
\\
&= -u(x)H(x)A_\nu(y)\del^\nu H(y)\del^\mu_x 
D(x-y),
\end{align*}
plus the analogous one in~$[T_1(x), Q_1(y)]$.  We are led to
the singular part
\begin{equation}
-2uH(A\.\del H)\dl(x-y).
\label{eq:some-trouble-four}
\end{equation}
The short rule here is $\del^\mu D\to2\dl$.

Next, we obtain
\begin{align*}
&\frac{im^2_H}m u(x)H(x)[\del^\mu B(x), B(y)]B(y)H(y)
\\
&= \frac{m^2_H}m u(x)H(x)B(y)H(y)\del^\mu_x 
D(x-y),
\end{align*}
leading to the singular part
\begin{equation}
\frac{2m^2_H}m uBH^2\dl(x-y).
\label{eq:some-trouble-six}
\end{equation}

\smallskip

From the last term in~$Q_1$, combining with the first term in~$T_1(y)$,
we obtain in all the singular part
\begin{equation}
2muB(A\.A)\dl(x-y).
\label{eq:some-trouble-seven}
\end{equation}

Combining both third terms, we consider
\begin{align*}
&-iu(x)B(x)[\del^\mu H(x), H(y)]A_\nu(y)\del^\nu B(y) \\
&= -u(x)B(x)A_\nu(y)\del^\nu B(y)\del^\mu_x D_{m_H}(x-y).
\end{align*}
We have in all the singular part:
\begin{equation}
-2uB(A\.\del B)\dl(x-y).
\label{eq:some-trouble-eight}
\end{equation}

Coming from respectively the third term in~$Q$ and the fifth term
in~$T_1$, there is the commutator
\begin{align*}
&\frac{-im_H^2}{2m}u(x)B(x)[\del^\mu H(x), H(y)]B^2(y)
\\
&= -\frac{m_H^2}m u(x)B(x)B^2(y) \del^\mu_xD_{m_H}(x-y).
\end{align*}
After collecting the similar term and taking the divergences, this
leads to
\begin{equation}
-\frac{m_H^2}{m^3}u B^3\dl(x-y).
\label{eq:some-trouble-ten}
\end{equation} 
Coming respectively from the third and sixth term, there is the
commutator
\begin{align*}
&3imeu(x)B(x)[\del^\mu H(x), H(y)]H^2(y)
\\
&= 3emu(x)B(x)H^2(y)\del^\mu_x D_{m_H}(x-y).
\end{align*}
After taking the divergences, this leads to a total singular part
\begin{equation}
6emuBH^2\dl(x-y).
\label{eq:some-trouble-eleven}
\end{equation}

\medskip

Next we list all possible normalization terms.  Among them, the two
first ones are coming from second-order tree graphs with two
derivatives on the inner line.  In other words, they come from
$s[T_1(x), T_1(y)]$.  Indeed, in this causal distribution, combining
the third terms in the expression of~$T_1$, there appears the term
\begin{align*}
& iA_\mu(x)H(x)[\del^\mu B(x), \del^\nu B(y)]A_\nu(y) H(y)
\\
&= A_\mu(x)H(x)A_\nu(y)H(y)\del_x^\mu\del_x^\nu D(x - y).
\end{align*}
This leads us to a normalization term $C_B(A\.A)H^2 \dl(x-y)$.  By the
same token, the reader may verify that combining the fourth terms in
the expression of~$T_1$ there appears the normalization term
$C_H(A\.A)B^2 \dl(x-y)$.

However, any term of the same form, compatible with Poincar\'e
covariance, discrete symmetries, ghost number and power counting
represents in principle a legitimate normalization.  Thus we introduce
the list of (re)normal\-ization terms we need:
\begin{align*}
N_2^1 &= C_B(A\.A)H^2\dl(x - y);
\\
N_2^2 &= C_H(A\.A)B^2\dl(x - y);
\\
N_2^3 &= -\frac{m_H^2}{4m^2}B^4\dl(x - y);
\\
N_2^4 &= \biggl(\frac{m_H^2}{m^2} + 3e\biggr)
B^2H^2\dl(x - y).
\end{align*}
In view of~\eqref{eq:dance-with-her} they generate new couplings.
There is also a $N_2^5$ term in~$H^4$, that we omit for now.  For
convenience, we have anticipated the coefficients in $N_2^3,N_2^4$,
which are of the second class.  The normalization terms amount to new
vertices with four external legs.  We compute the coboundaries:
\begin{align*}
sN_2^1 &= 2C_BH^2(A\.\del u)\dl(x - y);
\\
sN_2^2 &= 2C_H[B^2(A\.\del u) + muB(A\.A)]\dl(x - y);
\\
sN_2^3 &= -\frac{m_H^2}m uB^3\dl(x - y);
\\
sN_2^4 &= \Bigl(\,\frac{2m_H^2}m + 6em\Bigr) uBH^2\dl(x - y).
\end{align*}

The cancellation now is easy to obtain: let $C_B=C_H=1$.  This means
that we have only to worry about the first two terms
in~\eqref{eq:bitter-end} and similarly in~\eqref{eq:RIP}.  Now,
respectively the term~\eqref{eq:some-trouble-four} cancels the second
one in~\eqref{eq:bitter-end} and the
term~\eqref{eq:some-trouble-eight} cancels the second one
in~\eqref{eq:RIP}.  The two remaining terms in~\eqref{eq:bitter-end}
and~\eqref{eq:RIP}, together
with~\eqref{eq:some-trouble-six},~\eqref{eq:some-trouble-seven},
\eqref{eq:some-trouble-ten} and~\eqref{eq:some-trouble-eleven} are
exactly accounted for thanks to the normalization summands.

Therefore we have determined $T_1$ and $T_2$, except that $e$ still
remains indeterminate.  But please read on.

\subsection{Higher-order analysis}

For the higher-order analysis, it is convenient to have the expansion of the
inverse $\Sf$-matrix:
$$
\Sf^{-1}(g) =: 1 + \sum_1^\infty\frac{i^n}{n!}\int d^4x_1\dots\int
d^4x_n\, {\owl T}_n(x_1,\dots,x_n)\,g(x_1)\dots g(x_n).
$$
For
instance, the second order term ${\owl T}_2(x_1,x_2)$ in the expansion
of~$\Sf^{-1}(g)$ is given by
$$
{\owl T}_2(x_1,x_2) = -T_2(x_1,x_2) + T_1(x_1)T_1(x_2) +
T_1(x_2)T_1(x_1).
$$
Then, say,
\begin{align*}
A_3(x_1,x_2,x_3) &= {\owl T}_1(x_1)T_2(x_2,x_3) + {\owl
T}_1(x_2)T_2(x_1,x_3) + {\owl T}_2(x_1,x_2)T_1(x_3)
\nonumber \\
&+ T_3(x_1,x_2,x_3);
\nonumber \\
R_3(x_1,x_2,x_3) &= T_1(x_3){\owl T}_2(x_1,x_2) + T_2(x_1,x_3){\owl
T}_1(x_2) + T_2(x_2,x_3){\owl T}_2(x_1)
\nonumber \\
&+ T_3(x_1,x_2,x_3).
\end{align*}

\newpage
Just as before, $D_3:=R_3-A_3$ depends only on known quantities, is
causal in~$x_3$ and is gauge invariant.  Splitting it, we can
calculate $T_3$.  We refer to~\cite{Zurichneverdies} for the outcome
of the analysis in our case, which turns out to be quite simple.  The
the missing cubic term is given by:

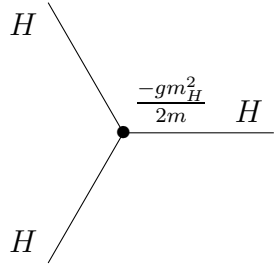
\begin{figure}[htb]
\centering
\begin{tikzpicture}
\draw (0,0) node {$\bullet$} node[above right] {$\frac{-gm_H^2}{2m}$}
   -- (2,0) node [above left] {$H$};
\draw (0,0) -- (120:2) node[below left] {$H$};
\draw (0,0) -- (240:2) node[above left] {$H$};
\end{tikzpicture}
\end{figure}

Also, it is seen that we need the new normalization term
$N_2^5=fH^4\dl$.  One finds $f=-m_H^2/4m^2$, and we are home.

\subsection{Summary of the abelian model}

Thus we write down the final (interaction) Lagrangian associated to
the abelian theory of the previous section.  There are two physical
fields $A^\mu,H$, of respective masses $m,m_H$, and an assortment of
ghosts $u,\ut,B$, which in our Feynman gauge all possess mass~$m$.  We
obtained six cubic couplings (proportional to~$g$) and five quartic
ones (proportional to~$g^2$).  It is remarkable that CGI generates the
latter from the former.  Only four terms out of the eleven involve
couplings exclusively among the physical fields.

\begin{align*}
\L_{\rm int}(x) &= gm(A\.  A)H - gm\ut uH + gB(A\.\del H)
\\
&- gH(A\.\del B) - \frac{gm_H^2}{2m}H^3 -
\frac{gm_H^2}{2m}B^2H
\\
&+ \frac{g^2}2(A\. A)H^2 + \frac{g^2}2(A\. A)B^2
\\
& - \frac{g^2m_H^2}{8m^2}H^4 - \frac{g^2m_H^2}{4m^2}H^2B^2 -
\frac{g^2m_H^2}{8m^2}B^4.
\end{align*}
With $C=1/m$ this tails down perfectly with~\eqref{eq:romeros-somos}
together
with~\eqref{eq:abyssus-abyssum-invocat},~\eqref{eq:dente-superbo}
and~\eqref{eq:hoc-erat-in-votis}.  By construction the total
Lagrangian is BRS invariant in the sense defined here.  (It has been
proved recently in a rigorous way~\cite{Michael05} in the EG framework
for interacting fields that ``classical'' BRS invariance implies gauge
invariance for all tree graphs at all orders.)

We exhibit the quartic interaction vertices graphically.

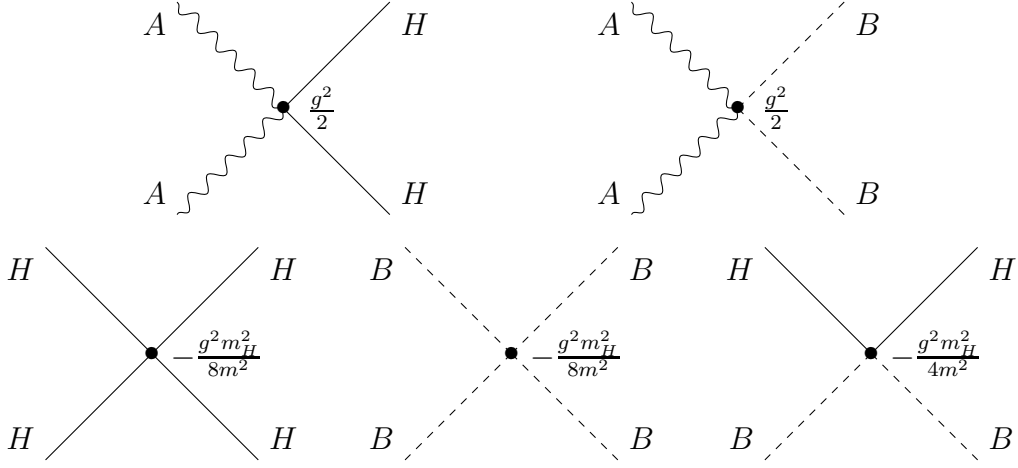
\begin{figure}[htb]
\centering
\begin{tikzpicture}
\draw (0,0) node {$\bullet$} node[right=5pt] {$\frac{g^2}2$}
   -- (45:2) node [below right] {$H$};
\draw[snake=snake] (0,0) -- (135:2) node [below left] {$A$};
\draw (0,0) -- (-45:2) node [above right] {$H$};
\draw[snake=snake] (0,0) -- (-135:2) node [above left] {$A$};
\end{tikzpicture}
\hspace{4pc}
\begin{tikzpicture}
\draw[dashed] (0,0) node {$\bullet$}
   node[right=5pt] {$\frac{g^2}2$}
   -- (45:2) node [below right] {$B$};
\draw[snake=snake] (0,0) -- (135:2) node [below left] {$A$};
\draw[dashed] (0,0) -- (-45:2) node [above right] {$B$};
\draw[snake=snake] (0,0) -- (-135:2) node [above left] {$A$};
\end{tikzpicture}
\end{figure}

\begin{figure}[htb]
\centering
\begin{tikzpicture}
\draw (0,0) node {$\bullet$} node[right=3pt] {$-\frac{g^2m_H^2}{8m^2}$}
   -- (45:2) node [below right] {$H$};
\draw (0,0) -- (135:2) node [below left] {$H$};
\draw (0,0) -- (-45:2) node [above right] {$H$};
\draw (0,0) -- (-135:2) node [above left] {$H$};
\end{tikzpicture}
\hspace{1pc}
\begin{tikzpicture}
\draw[dashed] (0,0) node {$\bullet$}
   node[right=3pt] {$-\frac{g^2m_H^2}{8m^2}$}
   -- (45:2) node [below right] {$B$};
\draw[dashed] (0,0) -- (135:2) node [below left] {$B$};
\draw[dashed] (0,0) -- (-45:2) node [above right] {$B$};
\draw[dashed] (0,0) -- (-135:2) node [above left] {$B$};
\end{tikzpicture}
\hspace{1pc}
\begin{tikzpicture}
\draw (0,0) node {$\bullet$} node[right=3pt] {$-\frac{g^2m_H^2}{4m^2}$}
   -- (45:2) node [below right] {$H$};
\draw (0,0) -- (135:2) node [below left] {$H$};
\draw[dashed] (0,0) -- (-45:2) node [above right] {$B$};
\draw[dashed] (0,0) -- (-135:2) node [above left] {$B$};
\end{tikzpicture}
\end{figure}

\smallskip

Notice that the purely scalar couplings are
\begin{align*}
&-g\frac{m_H^2}{2m}H(B^2 + H^2)
-g^2\frac{m_H^2}{8m^2}(B^2 + H^2)^2
\\
&= -\,\frac{g^2m_H^2}{8m^2}(B^2 + H^2)\big(B^2 + H^2 +
\frac{4m}{g}H\big).
\end{align*}
Performing now an asymptotic analysis (that is, taking the
St\"uckelberg field $B=0$) it becomes
$$
-\,\frac{g^2m_H^2}{8m^2}\big(H^4 + \frac{4m}{g}H^3\big).
$$

\section{Three MVBs}

Let us now seek all gauge theories with \textit{three} gauge fields.
The only interesting Lie algebra entering the game is
$$
\g = \mathfrak{su}(2);
$$
in this case obviously total antisymmetry implies the Jacobi identity.

The case $m_1=m_2=m_3=0$ is certainly possible, and then neither
scalar Higgs nor St\"uckelberg fields are necessary.

The simplest of the \textit{mass relations} we referred to in
Section~2 is the following: if $f_{abc}\ne0$ and $m_a=0$, then
necessarily $m_b=m_c$.  We see at once from this that if $m_1=0$ must
be $m_2=m_3$: the case $m_1=m_2=0,m_3\ne0$ is downright impossible.

The only other mass relation one needs to check to verify that models
with two or three MVBs and one Higgs-like field are correct in our
sense is
\begin{align}
4C^2m_b^2m_a^2 &= 2(m_a^2 + m_b^2) \sum_{d:m_d=0}\! (f_{abd})^2
\nn \\
&+ \sum_{k:m_k\neq 0}\!  \frac{(f_{abk})^2}{m_k^2} \bigl[ (m_a^2 +
m_b^2 + m_k^2)^2 - 4(m_a^2 m_b^2 + m_k^4) \bigr].
\label{eq:mother-lode}
\end{align}
With $C^{-1}=\pm m_2$, the model with the mass pattern $m_2=m_3\ne0,
m_1=0$ passes muster.

If we assume that all masses are different from zero, then necessarily
$m_1=m_2=m_3$.  Indeed, equation~\eqref{eq:mother-lode} implies
$$
4 m_a^2 m_b^2 m_c^2 C^2
= \bigl[
(m_a^2 + m_b^2 + m_c^2)^2 - 4(m_a^2 m_b^2 + m_c^4) \bigr]
$$
where $(a,b,c)$ is any permutation of $(1,2,3)$. Therefore,
$$
m_1^2 m_2^2 + m_3^4 = m_2^2 m_3^2 + m_1^4 = m_3^2 m_1^2 + m_2^4.
$$
This yields
\begin{align*}
(m_1^2 m_2^2 + m_3^4) - (m_2^2 m_3^2 + m_1^4)
&= (m_3^2 - m_1^2)(m_1^2 - m_2^2 + m_3^2) = 0,
\\
(m_2^2 m_3^2 + m_1^4) - (m_3^2 m_1^2 + m_2^4)
&= (m_1^2 - m_2^2)(m_2^2 - m_3^2 + m_1^2) = 0,
\end{align*}
whose only all-positive solution is $m_1 = m_2 = m_3 =: m$; and then
$4m^6 C^2 = m^4$ yields $C^{-1}=\pm2m$.

\smallskip

Physically, the two cases just examined correspond respectively to the
Georgi--Glashow model of electroweak interactions without neutral
currents; and to the~$\mathfrak{su}(2)$ Higgs--Kibble model.
Reference~\cite{CabezondelaSal} claims that more than one Higgs-like
particle for the $\mathfrak{su}(2)$ Higgs--Kibble model is not
allowed.  It is well known that the first mass pattern obtained here
is arrived at by SSB when the Higgs sector is chosen to be a $SU(2)$
isovector; and the second one when it is a complex doublet.  But in
our derivation SSB played no role.

\section{The Weinberg--Salam model within CGI}

Scharf and coworkers (see references in the introduction) followed a
``deductive'' approach to the SM, with the only assumption that $m_1$,
$m_2$, $m_3$ are all positive, plus existence of the photon, that is,
$m_4 = 0$.  There is no point in repeating that.  Suffice to say that
a structure constant like $f_{124}$ is found to be non-zero, thus
$m_1=m_2$; and also the mass constraints imply $m_3>m_1$.  Defining
\begin{equation*}
\cos \thW := m_1/m_3,
\end{equation*}
it is possible now to take for the non-zero structure constants
$$
|f_{123}| = \cos \thW \sepword{and} |f_{124}| = \sin \thW.
$$
With this, simply bringing~\eqref{eq:romeros-somos} together with
equations
with~\eqref{eq:abyssus-abyssum-invocat},~\eqref{eq:dente-superbo}
and~\eqref{eq:hoc-erat-in-votis}, one retrieves the boson part of the
SM Lagrangian, as given for example in~\cite{PapaTomate}.

Thus it appears that the ordinary version of the Higgs sector for the
gauge group $SU(2)\x U(1)\simeq U(2)$ is ``chosen'' by CGI. Of course,
one can argue for it from other considerations within the SSB
framework, or refer to experiment. We comment in the final discussion
on the problem of determining which patterns of broken symmetry are
allowed in CGI for general gauge~groups.

\subsection{Coupling to matter}

Things stay interesting when considering the fermion sector.  The
basic interaction between carriers and matter in a gauge theory is of
the form
$$
g(b^aA_{a_\mu}\ovl\psi\ga^\mu\psi + {b'}^aA_{a_\mu}
\ovl\psi\ga^\mu\ga^5\psi),
$$
with $\ovl\psi$ the Dirac adjoint spinor and $b,b'$ appropriate
coefficients.  In dealing with the SM our fermions are the known ones,
fulfilling as free fields the Dirac equation: we do not assume chiral
fermions \textit{ab initio}.  Their gauge variation is taken to be
zero.  Thus for the SM one makes the Ansatz
\begin{align}
T_1^F
&= b_1W_\mu^+\bar e\ga^\mu\nu + b'_1W_\mu^+\bar e\ga^\mu\ga^5\nu
+ b_2W_\mu^-\bar\nu\ga^\mu e + b'_2W_\mu^-\bar\nu\ga^\mu\ga^5 e
\nn \\
&\quad + b_3Z_\mu\bar e\ga^\mu e + b'_3Z_\mu\bar e\ga^\mu\ga^5 e +
b_4Z_\mu\bar\nu\ga^\mu\nu + b'_4Z_\mu\bar\nu\ga^\mu\ga^5\nu
\nn \\
&\quad + b_5A_\mu\bar e\ga^\mu e + b'_5A_\mu\bar e\ga^\mu\ga^5 e +
b_6A_\mu\bar\nu\ga^\mu\nu + b'_6A_\mu\bar\nu\ga^\mu\ga^5\nu
\nn \\
&\quad + c_1B^+\bar e\nu + c'_1B^+\bar e\ga^5\nu
+ c_2B^-\bar\nu e + c'_2B^-\bar\nu\ga^5 e
\nn \\
&\quad + c_3B_Z\bar e e + c'_3B_Z\bar e\ga^5 e +
c_4B_Z\bar\nu\nu + c'_4B_Z\bar\nu\ga^5\nu
\nn \\
&\quad + c_5H\bar\nu\nu + c'_5H\bar\nu\ga^5\nu + c_6H\bar e e +
c'_6H\bar e\ga^5 e.
\label{eq:canto-en-los-dientes}
\end{align}
Here~$e$ stands for an electron, muon or neutrino or a (suitable
combination of) quarks $d,s,b$; and $\nu$ for the neutrinos or the
quarks $u,c,t$; the charge difference is always minus one.  For
instance in the ``vertex'' $W_\mu^+\bar e\ga^\mu\nu$ a ``positron''
exchanges a $W^+$~boson and becomes a ``neutrino''.  Charge is
conserved in each term.

The method to determine the coefficients
in~\eqref{eq:canto-en-los-dientes} \textit{remains the same}; only, it
is simpler in practice.  We limit ourselves to a few remarks.  The
direct equation
$$
sT_1^F = i\del\.Q^F_1
$$
already allows to determine $c_1,c'_1,c_2,c'_2,c_3,c'_3,c_4,c'_4$, as
well as the vanishing of~$b'_5$ and~$b'_6$, assuming nonvanishing
fermion masses.  (For $\nu$ representing a true neutrino, we expect
the term with coefficient $b'_6$ to vanish anyway, since the photon
should not couple to uncharged particles.  The same is true
for~$b_6$.)  Thus the photon has no axial-vector couplings,
``because'' there is no St\"uckelberg field for it, that is, because
it is massless.  The reader will have no trouble in finding the
explicit form of $Q^F_1$, that can be checked
with~\cite[Eq.~4.7.4]{Zurichneverdies}.  At second order, one needs to
take into account the interplay of contractions between~$Q_1$
and~$T_1^F$, as well as the ``purely fermionic'' ones between~$Q^F_1$
and $T_1^F$.  There are no contractions between $Q^F_1$ and $T_1$,
since the former does not contain derivatives.  Also, \textit{no new
normalization terms} with fermionic fields may be forthcoming in
$sN_2$ or $\del_x\.N_{2/1}, \del_y\.N_{2/2}$, since a term
$\sim\vf_1\vf_2\bar\psi\psi\dl$ would be nonrenormalizable by power
counting: the only way to cancel local terms is that the coefficient
of every generated Wick monomial add up to zero.

At the end of the day, the physical Higgs couplings are proportional
to the mass, and \textit{chirality} of the interactions is a
\textit{consequence} of CGI~\cite{PGI-EW-I,PGI-EW-II}.  For leptons it
yields:
\begin{align*}
T_1^F
&= \frac1{2\sqrt2}W_\mu^+\bar e\ga^\mu(1 \pm \ga_5)\nu 
+ \frac1{2\sqrt2}W_\mu^-\bar\nu\ga^\mu(1 \pm \ga_5)e 
 + \frac1{4\cos\thW}Z_\mu\bar e\ga^\mu(1 \pm \ga_5)e
\\
&\quad
- \sin\thW\tan\thW Z_\mu\bar e\ga^\mu e
-\frac1{4\cos\thW}Z_\mu\bar\nu\ga^\mu(1 \pm \ga_5)\nu
+ \sin\thW A_\mu\bar e\ga^\mu e
\\
&\quad + i\frac{m_e - m_\nu}{2\sqrt{2}m_{\mathrm{W}}}B^+\bar e\nu
\pm i\frac{m_e + m_\nu}{2\sqrt{2}m_{\mathrm{W}}}B^+\bar e\ga^5\nu
-i\frac{m_e - m_\nu}{2\sqrt{2}m_{\mathrm{W}}}B^-\bar\nu e
\\
&\quad \pm i\frac{m_e + m_\nu}{2\sqrt{2}m_{\mathrm{W}}}B^-\bar\nu\ga^5 e
\pm i \frac{m_e}{2m_{\mathrm{W}}}B_Z\bar e e \pm i
\frac{m_\nu}{2m_{\mathrm{W}}}B_Z\bar e\ga^5
\\
&\quad + \frac{m_\nu}{2m_{\mathrm{W}}}H\bar\nu\nu +
\frac{m_e}{2m_{\mathrm{W}}}H\bar e e,
\end{align*}
as it should. In summary we have recovered the SM, with its rationale 
upside-down.

\section{Discussion}

People define $e=g\sin\thW;\,g'=g\tan\thW$.  Therefore,
$$
\sec\thW = \frac{\sqrt{g^2 + g'{}^2}}{g},
$$
and selecting the chiral projector and apart from the standard
factors, the effective coupling of the term in $W_\mu^+\bar
e\ga^\mu\nu$ and conjugate is~$g$; that of the term $A_\mu\bar
e\ga^\mu e$ is~$e$; that of the term $Z_\mu\bar\nu\ga^\mu\nu$ is
$-\sqrt{g^2 + {g'}^2}/2g$; and so on.  Thus one can artfully write
things as if $g,g'$ are two different coupling constant associated to
the emerging representation of the gauge group.  But we have seen that
the coefficients come from the pattern of masses, which in our
viewpoint is fixed by nature.  In order to bring home the point, let
us make the \textit{Gedankenexperiment} of building the SM from the
Georgi--Glashow model, by adding a vector boson, sitting on an
invariant abelian subgroup.  Implicitly we allow for two different
coupling constants (plus mixing of the old photon and the new MVB).
But in that case there is no reason for $m_Z>m_W$.  It is more natural
to assume that the SM stems from the Higgs--Kibble model, keeping one
coupling constant, whereby two of the three masses are moderately
``pulled down'' by mixing with the new photon.  This goes to the heart
of the experimental situation; other weak isospin values do not enter
the game.  In other words, no support comes from our quarter to the
idea that the SM as it stands is ``imperfectly unified''.  The
argument is bolstered by the fact that the true group
of the electroweak interaction is~$U(2)$, not $SU(2)\x U(1)$.%
\footnote{To our knowledge, this was noticed first in~\cite{Florian}.}

In usual presentations of the SM the $U(2)$ symmetry is said to be
``broken'', among other reasons, because there is only one conserved
quantity, electric charge, instead of four.  In CGI the interaction
appears to respect the $U(2)$ symmetry.  But of course symmetry is
broken already at the level of the free Lagrangian, due to different
masses (the residual symmetry $m_1=m_2$ goes in hand with electric
charge conservation).  This is to say that not all bases of the Lie
algebra are equivalent, since there is a natural basis dictated by the
pattern of masses.  The role of the mass constraints is precisely to
pick out this basis.%
\footnote{In the early seventies, speculations on fermionic patterns
of masses from SSB were rife ---see for instance~\cite{GG}.  They have
been since all but abandonded.  They might perhaps become a
respectable subject of study again in CGI. By now we may only say that
differences between fermion masses are related to differences between
boson masses in that (disregarding family mixing) models in which all
bosons share the same mass would entail identity of all fermion masses
as well.}

\smallskip

Let us recapitulate. CGI is a tool for the actual construction of
Lagrangians. We limited ourselves to polynomial couplings. At first
order in the coupling constant, CGI fixes some of the couplings of the
vector bosons and ghost (fermionic and bosonic) fields. At second
order, it requires additional quartic couplings, as well as some extra
ingredient, which here is made out of physical scalars or Higgs-like
fields. Third-order invariance goes on to fix the remaining couplings
of the Higgs-like fields. One obtains in that way potentials of the
symmetry-breaking kind, although SSB does not enter the picture.
Ockham's razor, already invoked in Section~2 in relation with the
number of Higgs fields, seems even more pertinent here.

\smallskip

On the historical side, it is difficult to imagine the development of
electroweak unification during the sixties without the SSB crutch.
Massive vector bosons were beyond the pale then.  The only
contemporaneous article (still instructive today) I know of, willing
and eager to start from them as fundamental entities
is~\cite{Samizdat}; it did not have enough impact.  Around ten years
later, after the invention of SSB, cogent arguments based on tree
unitarity ---see~\cite{CornwallLevinTiktopoulosPRD-10} and references
therein--- weighed in favour of the phenomenological outcome of gauge
theories with broken symmetry, plus abelian mass terms for invariant
abelian subgroups.  This is basically what CGI constructs.

Since they lead to the same phenomenological Lagrangian, there seems
to be no way as yet ---within ordinary particle physics, at least---
to distinguish between the~SM as presented in textbooks and its causal
version.  This is good, because it shows that CGI is solidly anchored
in physics.

It is also bad: ``a difference, to be a difference, has to make a
difference''. Still, a constructive CGI program was in principle
attractive because the apparent severity of the constraints on the
masses of the gauge fields. Ambauen and Scharf~\cite{FortunaJuvet}
argued that the $SU(5)$ grand unification model by Georgi and Glashow
with its standard pattern of Higgs fields~\cite[Chap.~18]{Georgi99},
is not causally gauge invariant; and the situation in this respect for
a while was murky. However, a systematic comparison between CGI and
the general theory of broken local symmetries~\cite{Ling-FongLi} has
been performed recently~\cite{VierMaenner}, and the contention
of~\cite{FortunaJuvet} that there might be contradiction between
causal gauge invariance and some grand unified models has been laid to
rest.

\subsection*{Acknowledgments}

I am most grateful for discussions to Luis J. Boya, Florian Scheck and
Joseph~C. V\'arilly. Special thanks are due to Michael D\"utsch, who
patiently explained to me aspects of the gauge principle according to
the Z\"urich school. I~acknowledge support from CICyT, Spain, through
grant~FIS2005--02309.

\end{document}